\documentclass[preprint,12pt,authoryear]{elsarticle}
\usepackage{amsmath,amsfonts}
\usepackage{algorithm}
\usepackage{algpseudocode}
\usepackage{array}
\usepackage[caption=false,font=normalsize,labelfont=sf,textfont=sf]{subfig}
\usepackage{textcomp}
\usepackage{stfloats}
\usepackage{url}
\usepackage{verbatim}
\usepackage{graphicx}
\usepackage{adjustbox}
\usepackage{booktabs}
\usepackage{pifont}

\usepackage{amssymb}
\usepackage{amsmath}

\journal{Nuclear Physics B}

\begin{document}

\begin{frontmatter}



\title{Multi-Granularity Position Embedding of Graphs via Granular-Ball for Link Prediction} 


\author[2]{Sen Zhao}
\author[1]{Cheng Liu}
\author[1]{Shuyin Xia*}
\author[1]{Zhiyuan Liu}
\author[3]{Liu Yi}
\author[3]{Yi Wang}
\author[3]{Wei Wang}

\affiliation[1]{organization={Chongqing Key Laboratory of Computational Intelligence, Key Laboratory of Big Data Intelligent Computing, Key Laboratory of Cyberspace Big Data Intelligent Security, Ministry of Education, School of Computer Science and Technology, Chongqing University of Posts and Telecommunications}, 
 city={Chongqing}, 
 postcode={400065}, 
 country={China}}

\affiliation[2]{organization={Chongqing Key Laboratory of Computational Intelligence, Key Laboratory of Big Data Intelligent Computing, Chongqing University of Posts and Telecommunications},
 city={Chongqing}, 
 postcode={400065}, 
 country={China}}

 \affiliation[3]{organization={Chongqing Ant ConsumerFinance Co,. Ltd, Ant Group},
 city={Chongqing}, 
 postcode={400065}, 
 country={China}}
            
\begin{abstract}
Link prediction aims to identify potential or future connections within a given graph structure. Position information is essential for link prediction, as it distinguishes homogeneous nodes through their relative relationships, facilitating the accurate capture of structural patterns and implicit connections. Previous studies derive node positional information as distances to single-granularity landmarks, defined as the centers of homophilic regions, while neglecting the multi-granularity nature of homophilic structures and their hierarchical interrelations. We propose the Multi-Granularity Position Embedding of Graphs via Granular-Ball for Link Prediction (MGLP) method to obtain multi-granularity position embedding of graphs. Specifically, MGLP introduces an Adaptive Granular-Ball Graph Refinement mechanism to adaptively refine the graph into homophilic subdomains with optimal levels of granularity. The central nodes within subdomains are treated as landmarks, which form a Hierarchical Central Graph. Moreover, a novel Multi-granularity Hierarchical Distance encoding mechanism is proposed to capture both the homophilic structures within a graph and their hierarchical correlations, improving the discriminative power of nodes. Experimental results demonstrate that the multi-granularity position embedding generated by our method exhibits excellent performance and strong competitiveness compared to baseline algorithms for link prediction. Our codes are available in https://anonymous.4open.science/r/MGLP-D3C5/.
\end{abstract}




\end{frontmatter}
\section{Introduction}
Link prediction seeks to uncover potential links or future connections that are not yet observed within a given graph structure. By identifying implicit relationships, link prediction supports various applications, including friend recommendation, product recommendation in e-commerce \cite{wang2019knowledge,you2019hierarchical}, social networks \cite{adamic2003friends}, knowledge graph completion \cite{nickel2015review},  protein interaction analysis \cite{airoldi2008mixed}, and beyond. 

Early link prediction methods rely on heuristics to quantify node similarity and predict link likelihood \cite{barabasi1999emergence}. However, their manual design for specific network structures limits generalizability and cross-domain applicability. To address this limitation, Graph Neural Networks (GNNs) have been proposed as an effective solution for learning structured node representations through neighborhood aggregation \cite{gilmer2017neural,kipf2017semisupervised,hamilton2017inductive,velickovic2018graph}. However, standard GNNs fall short of distinguishing homogeneous nodes, which adversely affects their performance in link prediction tasks. 
Position-embedding methods have been proposed to address this issue \cite{you2019position,dwivedi2020benchmarking,kreuzer2021rethinking,wang2022equivariant}, leveraging nodes' positional information as a feature based on their relative distances to other nodes. This positional information is crucial for link prediction, enabling the differentiation of homophilic nodes and capturing relationships between node pairs effectively. HPLC \cite{kim2024hierarchical} enhances the scalability of position-embedding methods by introducing landmarks to derive node positional information relative to them instead of computing pairwise distances for all nodes, reducing computational complexity while preserving accuracy. However, this approach relies solely on unit-grained landmarks, failing to account for the multi-granularity of homophilic structures within the graph and their hierarchical correlations.
\begin{figure}[!t]
    \includegraphics[width=1\linewidth]{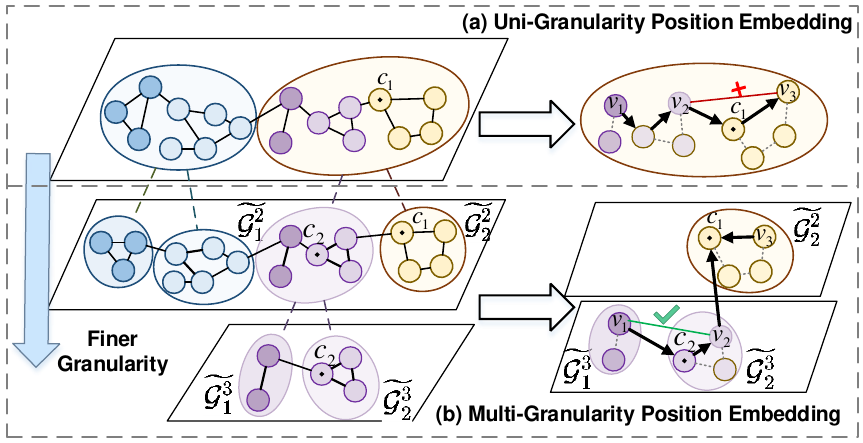}
    \caption{The comparison of Uni-Granularity and Multi-Granularity Position Embedding.}
    \label{fig:motivate}
\end{figure}

In real-world link prediction scenarios, graphs are inherently composed of homogeneous subdomains that exhibit varying levels of granularity, where nodes within each subregion demonstrate relatively stronger homophilic properties. As illustrated in Fig.~\ref{fig:motivate}, the graph as a whole can be considered the coarsest granularity of homophilic structure, with progressively finer-grained levels of structure represented from top to bottom. Each subdomain aligns optimally with a specific level of granularity.
 For the example in Fig.~\ref{fig:motivate} (b), $\widetilde{\mathcal{G}_1^2}$ is at the appropriate granularity, while $\widetilde{\mathcal{G}_2^2}$ should
 be further decomposed into $\widetilde{\mathcal{G}_1^3}$ and $\widetilde{\mathcal{G}_2^3}$ at a finer granularity. 
These varying levels of granularity naturally organize into a tree-like structure, where the closeness of nodes within the tree, defined by the depth of their nearest common ancestor relative to the nodes, reflects the strength of their correlation and homophilic properties. As shown in Fig.~\ref{fig:motivate}(b), the finer-grained landmark $c_2$ is more appropriate for estimating the homophily between nodes $v_1$ and $v_2$ compared to the coarser-grained landmark $c_1$. Additionally, the significant distance between $v_3$ and $c_1$ in the tree indicates their lack of relevance.
 Disregarding the multi-granularity of homophilic structures within the graph and their hierarchical correlations (as illustrated in Fig.~\ref{fig:motivate} a) may lead to the selection of landmarks at an unsuitable granularity level, resulting in inaccurate homophily estimations between nodes.

 Actually, it is non-trivial to model the multi-granularity distribution of homophily within the graph for link prediction, due to three challenges: 1) Homophilic structure refinement: 
Homophilic structures are formed by nodes with high connectivity and strong feature similarity. These structures should be modeled at the appropriate level of granularity, ensuring that neither a coarser-grained parent graph nor a finer-grained subgraph exhibits a higher homophily ratio than the level at which the structure is being analyzed.
Building on the strengths of granular-ball computing \cite{xia2019granular,xie2023efficient} in modeling multi-granular characteristics for scatter data, we further explore its application in graph refinement and introduce an Adaptive Granular-Ball graph refinement mechanism to capture homophilic structures within the graph.
2) Relative position encoding among landmarks: Representing the distance between nodes through landmarks is highly beneficial for improving the accuracy of estimating node position. Therefore, a more comprehensive landmark representation is required to provide a more efficient estimation of node position.
3) Position information encoding: The relative distance of a node to others is essential for distinguishing homogeneous nodes. Measuring these distances requires considering homophilic structures and their hierarchical correlations across different granularities. Thus, an effective position encoding mechanism is needed to capture multi-granularity relationships and provide a refined representation of node positions within the graph.

To address these challenges, we propose a novel method called \textbf{M}ulti-\textbf{G}ranularity Position Embedding of Graphs via Granular-Ball for \textbf{L}ink \textbf{P}rediction (MGLP). MGLP introduces an Adaptive Granular-Ball Graph Refinement mechanism that iteratively decomposes the graph into homophilic subdomains, ensuring that each subdomain is refined to the optimal level of granularity. Subsequently, the granular-ball refinement process is represented within the \textbf{H}ierarchical \textbf{C}entral \textbf{G}raph (HCG) which uses the initial granular-ball central nodes as the root of the tree and diffuses outward in a tree-like structure, with the hierarchical relationships in the tree reflecting the interconnections between landmarks. Moreover, a novel distance measurement mechanism called \textbf{M}ulti-\textbf{G}ranularity \textbf{H}ierarchical \textbf{D}istance (MGHD) is designed to capture both the homophilic structures within a graph and their hierarchical correlations across different granularities. MGHD employs the distances between nodes along tree-structured paths as weights, incorporating these weights to evaluate the relationship between nodes and landmarks. Extensive experiments conducted on six real-world datasets demonstrate that MGLP consistently outperforms state-of-the-art methods. 

Our contributions are summarized as follows:
\begin{itemize}
    \item We emphasize the importance of multi-granularity of homophilic structures within the graph and their hierarchical correlations for link prediction.
    \item We propose the MGLP method, which adaptively refines the graph into homophilic structures at optimal granularity and we utilize HCG to encode the relative positions among landmarks. Furthermore, we introduce MGHD to incorporate homophilic structures and their hierarchical correlations across multiple granularities.
    \item We conducted extensive experiments to demonstrate that the MGLP method is applicable to various types of datasets and consistently delivers outstanding performance.
\end{itemize}

\section{Related Work}
\subsection{Link Prediction}
Early research on link prediction primarily employed heuristic methods derived from social network analysis \cite{lu2011link}. Among these, the homophily mechanism \cite{mcpherson2001birds} suggests that "similar" nodes are more likely to connect. Most heuristics are based on measures of connectivity. Connectivity-based heuristics, such as the Adamić–Adar index (AA) \cite{adamic2003friends}, measure node similarity by counting shared neighbors, often resulting in predicted graphs with numerous triangular structures. AA primarily considers paths of length two, while other heuristics consider longer paths. For example, the Katz index \cite{katz1953new} computes a weighted sum of walks of all lengths between nodes. GNN methods for link prediction demonstrate the power of GNNs in learning effective graph representations. SEAL \cite{zhang2018link} introduces an $h$-hop enclosing subgraph centered on two target nodes, reformulating link prediction as a graph classification task based on the subgraph's structural topology.

Poistion-embedding methods for link prediction leverage both graph structure and node positional information. A notable example is the Graph Autoencoder (GAE) \cite{DBLP:journals/corr/KipfW16}, which uses a GCN to compute node representations and predicts links with a sigmoid function applied to node pair representations. The Variational Graph Autoencoder (VGAE) \cite{kipf2016variational} extends GAE by using two GCNs to learn the mean and variance of node embeddings, performing the link prediction task with the mean and variance of node pairs' embeddings. ARGE \cite{DBLP:journals/corr/abs-1802-04407} enhances GAE with adversarial regularization to align node embeddings with a prior distribution. S-VAE \cite{davidson2022hypersphericalvariationalautoencoders} replaces the Gaussian distribution in VGAE with a von Mises-Fisher distribution to model hyperspherical latent structures. P-GNN \cite{you2019position} aggregates messages from a subset of anchor nodes to capture positional information for link prediction. In knowledge graph completion, R-GCN \cite{schlichtkrull2018modeling} assigns different weights to relation types during message passing, while SACN \cite{shang2019end} performs message passing for each relation type and analyzes relations using weighted node embeddings. HPLC \cite{kim2024hierarchical} enhances the scalability of position-embedding methods by introducing the concept of landmarks, enabling the derivation of nodes' positional information relative to these landmarks rather than relying on pairwise distance calculations for all nodes.

\section{Methodology}
\begin{figure}[t]
\centering
    \includegraphics[width=1\textwidth]{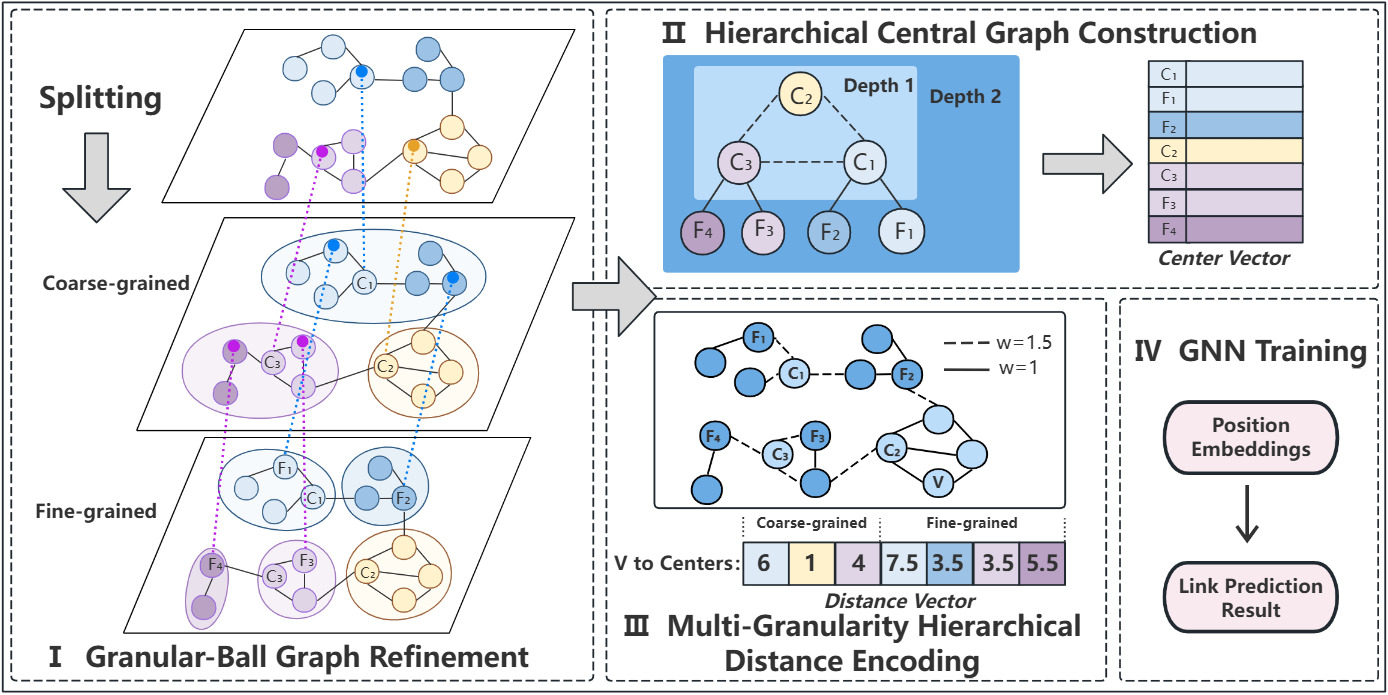}
    \caption{Overview of MGLP. \textbf{\MakeUppercase{\romannumeral 1}}. Granularizing the original graph, select the central node of each granular-ball.
    \textbf{\MakeUppercase{\romannumeral 2}}. Construct the HCG and compute center vectors. \textbf{\MakeUppercase{\romannumeral 3}}. Using MGHD and the granularization results to compute the distance vectors, with the node's hierarchical level represented by the color in the HCG. \textbf{\MakeUppercase{\romannumeral 4}}. Concatenate center vectors and distance vectors to form node position embeddings, and put these embeddings into GNN models. }
    \label{fig:model}
\end{figure}

As depicted in Figure \ref{fig:model}, our framework consists of four phases: (1) Granular-Ball Graph Refinement, which partitions the graph into granular balls of varying granularity and generates a tree based on the hierarchical structure when splitting; (2) Hierarchical Central Graph Construction, where the central nodes of granular balls are used to form a hierarchical central graph to encode relative positions among landmarks; (3) Multi-Granularity Hierarchical Distance Encoding, which captures both the homophilic structures within a graph and their hierarchical correlations across different granularities; and (4) GNN Training with Node Embeddings, combining center and distance vectors to generate node embeddings, which are subsequently utilized in GNN to predict link probabilities. We start this section by introducing important notations and definitions.

\subsection{Notations and Definitions}
An undirected graph is denoted as $\mathcal{G = (V, E, \mathbf{X}, \mathbf{A}),}$ where $\mathcal{V}$ represents the set of nodes with $\mathcal{|V| = N}$ indicating the total number of nodes, $\mathcal{E}$ represents the set of edges with $\mathcal{|E| = M}$ denoting the total number of edges, $\mathbf{X}$ is the feature matrix associated with the nodes, and $\mathbf{A}$ is the adjacency matrix of the graph.

\textbf{Definition 1 Granular-Ball on Graph:} A granular-ball, denoted as $\widetilde{\mathcal{G}}$, is defined as a connected subgraph of $\mathcal{G}$ and is formally expressed as $\mathcal{\widetilde{G} = (V}_{\widetilde{G}}, \mathcal{E}_{\widetilde{G}}, \mathbf{X}_{\widetilde{G}}, \mathbf{A}_{\widetilde{G}})$. Our approach refines the $\mathcal{G}$ into multiple regions as:
\begin{equation}
    \mathcal{G} = \{\widetilde{\mathcal{G}}_1, \widetilde{\mathcal{G}}_2, \ldots, \widetilde{\mathcal{G}}_n\}.
\end{equation}

\textbf{Definition 2 Granular-Ball Quality:} The quality $Q$ of a granular-ball $\widetilde{\mathcal{G}}$ is defined as the average connectivity of nodes within the granular-ball, which is calculated based on the number of edges and nodes in the granular-ball. The formula is as follows:
\begin{equation}
    \mathcal{Q(\widetilde{\mathcal{G}})} = \frac{2 \times \mathcal{|E}_{\widetilde{\mathcal{G}}}|}{\mathcal{|V}_{\widetilde{\mathcal{G}}}|},
\label{eq:Q}
\end{equation}
where $\mathcal{|E}_{\widetilde{\mathcal{G}}}|$ represents the number of edges between nodes within the granular-ball, and $\mathcal{|V}_{\widetilde{\mathcal{G}}}|$ represents the number of nodes in the granular-ball. 

\subsection{Granular-Ball Graph Refinement}\label{AA}
This phase can be further divided into coarse and fine-grained splitting, and the construction of the tree is carried out progressively during the splitting process. The initial step constructs the entire dataset into an original graph $\mathcal{G = (V, E, \mathbf{X}, \mathbf{A})}$. $\mathcal{G}$ is refined through a progressive splitting process, resulting in finer-grained granular-balls at each step. At refinement step $t$, the granular-ball $\widetilde{\mathcal{G}^{t}}$ is divided into a set of finer-grained granular-balls, denoted as $\{\widetilde{\mathcal{G}_1^{t+1}}, \widetilde{\mathcal{G}_2^{t+1}}, \ldots, \widetilde{\mathcal{G}_n^{t+1}}\}$. To represent this hierarchical decomposition to obtain center vectors and distance vectors, a tree $\mathcal{T}$ is constructed, where each node $\mathcal{T}^{t}_{i} = \{\widetilde{\mathcal{G}^{t}_{i}}, {c}^{t}_{i}, \mathcal{I}^{t}_{i}, \mathcal{P}^{t}_{i}\}$ means the corresponding granular-ball $\widetilde{\mathcal{G}^{t}_{i}}$, its central node ${c}^{t}_{i}$, its identifier $\mathcal{I}^{t}_{i}$ and the identifier of its parent granular-ball $\mathcal{P}^{t}_{i}$.

\textbf{Coarse-grained splitting}: The coarse-grained splitting phase involves dividing $\mathcal{G}$ into initial granular balls of varying sizes. During this process, the top $\mathcal{\alpha = \sqrt{N}}$ nodes with the highest degrees are selected from the node-set as the central nodes of the initial granular balls:
\begin{equation}
    \mathcal{C} = \{c_0, c_1, \ldots, c_{\alpha}\},
\end{equation}
where $\mathcal{C}$ is the set of central nodes and $\mathcal{N}$ is the total number of nodes. The value of $\mathcal{\sqrt{N}}$ is an empirical choice, drawing on previous work \cite{xie2020new}. This approach has been demonstrated to effectively balance computational complexity and accuracy. All nodes, except the selected central nodes, are assigned to central nodes as follows:
\begin{equation}
    \text{Assign}(v) = \{\widetilde{\mathcal{G}}_i \mid \min d(v, c_i), \, c_i \in \mathcal{C}\},
    \label{eq:Assign}
\end{equation}
where Assign(v) denotes that the node $v$ is assigned to the granular-ball $\widetilde{\mathcal{G}}_i$ and $d(v, c_i)$ is the shortest path distance between the node $v$ and the central node $c_i$ on $\mathcal{G}$. The initial granular balls are represented as:
\begin{equation}
    \mathcal{\widetilde{G_{\text{init}}}} = \{\widetilde{\mathcal{G}_1^{0}}, \widetilde{\mathcal{G}_2^{0}}, \ldots, \widetilde{\mathcal{G}_\alpha^{0}}\},
\end{equation}
The initial granular balls generated based on highly central nodes ensure strong connectivity, facilitating effective coverage of the graph. Each initial granular-ball ${\widetilde{\mathcal{G}_{i}^{0}}}\mathcal(\text{1} \leq {i} \leq \alpha)$ in $\mathcal{\widetilde{G_{\text{init}}}}$ undergoes independent fine-grained splitting. 

\textbf{Fine-grained splitting}: During this phase, the central node is selected as the node with the highest degree, excluding the node that has been selected as the central node to avoid splitting into an overly large granular-ball and an overly small granular-ball. The fine-grained splitting is performed through iterative binary splits. 

At the $t$-th step of splitting phase, given a granular-ball $\widetilde{\mathcal{G}^{t}_{i}}$, two nodes $c_1$ and $c_2$ with the highest degrees among the unselected nodes are chosen as the central nodes for the current partition to split $\widetilde{\mathcal{G}^{t}_{i}}$ into two finer-grained granular balls $\widetilde{\mathcal{G}^{t+1}_{1}}$ and $\widetilde{\mathcal{G}^{t+1}_{2}}$. Subsequently, all nodes in $\widetilde{\mathcal{G}^{t}_{i}}$ are assigned to the central node $c_1$ or $c_2$ using Eq.\ref{eq:Assign}. 

Once all nodes in $\widetilde{\mathcal{G}^{t}_{i}}$ have been assigned, the quality of $\widetilde{\mathcal{G}^{t+1}_{1}}$ and $\widetilde{\mathcal{G}^{t+1}_{2}}$ are computed using Eq.\ref{eq:Q}. A split is retained only if the resulting granular balls satisfy the following condition:
\begin{equation}
    {Q(\widetilde{\mathcal{G}^{t}_{i}})} < \max({Q(\widetilde{\mathcal{G}^{t+1}_{1}})}, {Q(\widetilde{\mathcal{G}^{t+1}_{2}})}),
\end{equation}
where this condition requires that at least one of the child granular balls has a higher quality than the parent granular-ball. This ensures that granular balls are refined in the direction of increasing quality.

With the completion of this phase, each granular-ball in $\mathcal{\widetilde{G_{\text{init}}}}$ is adaptively and iteratively refined, resulting in a set of ${\{\widetilde{\mathcal{G}_{i}}\}}$ with granular balls of different granularities and a tree $\mathcal{T}$ representing the hierarchical decomposition. 

\subsection{Hierarchical Central Graph Construction}

After obtaining ${\{\widetilde{\mathcal{G}_{i}}\}}$ and $\mathcal{T}$, we consider how to leverage them to construct the HCG. Our objective is to encode relative positions among landmarks to derive both global and local relationships of nodes.

We consider the graph consisting only of landmarks, which we refer to as central nodes. The central nodes of $\mathcal{\widetilde{G_{\text{init}}}}$ from the coarse-grained splitting phase are used to construct a fully connected graph. Subsequently, based on the refinement levels of each initial granular-ball and $\mathcal{T}$, we perform the edge addition operation on the connected graph as follows:
\begin{equation}
    \mathcal{E}^{\text{add}} = \{(c_j, c_k) \mid j=\mathcal{I}_{i},k=\mathcal{P}_{i}, \mathcal{T}_{i}\in \mathcal{T}\},
\end{equation}
where $\mathcal{E}^{\text{add}}$ represents edge addition operation, $\mathcal{I}_{i}$ is the identifier of granular-ball $\widetilde{\mathcal{G}_{i}}$, and $\mathcal{P}_{i}$ is the identifier of the parent granular-ball of $\widetilde{\mathcal{G}_{i}}$. Then tree-like structures are extended outward to form a multi-granularity granular-ball distribution graph, referred to as the HCG. The edge weights in the HCG, connecting two granular-ball central nodes $u$ and $v$, are defined as 
\begin{equation}
e_{uv} = \exp\left(-\frac{d(u, v)^2}{T}\right),
\end{equation}
where $T$ is the normalizing parameter of the heat kernel and $d(u, v)$ is the shortest path between $u$ and $v$ on $\mathcal{G}$.
Subsequently, we utilize the HCG to generate the multi-granularity center vector (\emph{CV}). 

\textbf{Center Vector}  For nodes belonging to the same granular-ball ${\widetilde{\mathcal{G}}}$, identical identity information is assigned to enhance consistency. This identity information is derived from the HCG, leveraging the relative positional relationships among granular-ball central nodes. Consequently, nodes within the same granular-ball ${\widetilde{\mathcal{G}}}$ share identical information, while those residing in the same coarse-grained granular-ball exhibit similar characteristics.

In this work, we introduce the HCG and utilize the graph Laplacian \cite{belkin2003laplacian} to encode the relative positional relationships among granular-ball central nodes. Let $\hat{A} \in \mathbb{R}^{K \times K}$ denote the weighted adjacency matrix of the central graph. The normalized graph Laplacian is computed as 
\begin{equation}
L = I - \Delta^{-\frac{1}{2}} \hat{A} \Delta^{-\frac{1}{2}},
\end{equation}
where $I$ is the identity matrix and $\Delta$ is the degree matrix with diagonal entries $\Delta_{ii} = \sum_{j} \hat{A}_{ij}$. The eigenvectors corresponding to the Laplacian $L$ are extracted and utilized as center vectors (\emph{CVs}), encoding the positional information of the granular-ball central nodes.

\subsection{Multi-Granularity Hierarchical Distance Encoding}

At this phase, we propose a novel distance computation paradigm MGHD that leverages ${\{\widetilde{\mathcal{G}_{i}}\}}$ and ${\mathcal{T}}$ to measure the distance between nodes. This method is specifically designed to simultaneously reflect the structural relationships in the original graph ${\mathcal{G}}$ and the hierarchical relationships in ${\mathcal{T}}$.

A limited subset of representative nodes, termed granular-ball central nodes, is selected to effectively capture the graph's structure. These nodes are represented as $c_1, c_2, \dots, c_K \in \mathcal{V}$, where $K$ indicates the total number of ${\{\widetilde{\mathcal{G}_{i}}\}}$. For a given node $u$, we define a $K$-dimensional tuple representing the distances from $u$ to each center $c_i$ ($i = 1, 2, \dots, K$). This tuple is referred to as distance vector (\emph{DV}):
\begin{equation}
	DV(u) = \left( d(u,c_1),d(u,c_2),\cdots,d(u, c_K) \right)
\end{equation}
For the distance, the metric is defined as the weighted sum of distances between nodes. Given a graph $\mathcal{G = (V, E),}$ where $\mathcal{V}$ and $\mathcal{E}$ represent the set of nodes and edges. Assume that the shortest path from node $u$ to node $v$ is $\mathcal{P}{ath} = (u,x_1,x_2,...,v)$, then the distance between two nodes $u$ and $v$, denoted as $d(u,v)$, is computed as follows:
\begin{equation}
	d(u, v) = \sum_{i=0}^{l} w(x_i, x_{i+1}),
\end{equation}
where $w(x_i, x_{i+1})$ is the weight of edge $(x_i, x_{i+1})$ and $l$ is the number of edges in the $\mathcal{P}{ath}$. The $w(x_i, x_{i+1})$ is computed as follows:
\begin{equation}
\begin{split}
w(x_i, x_{i+1}) &= 1 - \frac{1}{{Depth}_{max}} + \\
&\quad \frac{1}{{Depth}_{max} - |{Depth}(x_i) - {Depth}(x_{i+1})|}
\end{split},
\end{equation}
where $Depth(x_i)$ is the depth of node $x_i$ on $\mathcal{T}$ and $Depth_{max}$ means the maximum depth of $\mathcal{T}$.
By weighting distances at each hierarchical level, the metric represents both the homophilic structures within a graph and their hierarchical correlations across different granularities. 

\emph{CVs} and \emph{DVs} are jointly processed through a position encoder to generate node position embedding. 

\subsection{GNN Training with Node Embeddings}

The GNN model used in this study builds upon traditional GNNs by integrating local features, positional encodings, and the Jumping Knowledge (JK) \cite{xu2018representation} mechanism. Utilizing node features, node positional embedding, and the adja cency matrix to enhance the geometric representation capability of the features.

A three-layer fully connected network is employed to update local node features, enabling the capture of local node characteristics. The update for each node feature $\mathbf{h}_v$ is defined as:
\begin{equation}
\mathbf{h}_v^{(t+1)} = \sigma\left(\mathbf{W}^{(t)} \mathbf{h}_v^{(t)} + \mathbf{b}^{(t)}\right),
\end{equation}
where $\mathbf{W}^{(t)}$ and $\mathbf{b}^{(t)}$ are the weight matrix and bias for the $t$-th layer, and $\sigma$ is an activation function, such as ReLU.

The model aggregates neighborhood information using graph convolution operations. For a node $v$, its neighborhood aggregation is expressed as:
\begin{equation}
\mathbf{h}_v^{(t)} = \Phi\left(\{\mathbf{h}_u^{(t)} : u \in Neighbors(v)\}\right),
\end{equation}
where $\Phi(\cdot)$ is the neighborhood aggregation function, and $Neighbors(v)$ denotes the set of neighbors of node $v$. Depending on the architecture, $\Phi(\cdot)$ can represent mean aggregation (GCN), attention-based weighted sum (GAT), or sampling-based aggregation (GraphSAGE).

To address the over-smoothing problem, the model employs the JK mechanism to integrate node representations from multiple layers. The final representation of node $v$ is given by:
\begin{equation}
\mathbf{h}_v^{(JK)} = \Psi\left(\{\mathbf{h}_v^{(l)} : l = 1, \dots, L\}\right),
\end{equation}
where $\Psi(\cdot)$ is the layer combination function, which can be mean, max, concatenation, or LSTM-based combination.

Using the node position encoding, the model predicts links between node pairs. For a given pair of nodes $(v, u)$, the link score is computed as:
\begin{equation}
s_{vu} = \sigma\left(\mathbf{h}_v \cdot \mathbf{h}_u\right),
\end{equation}
where $\sigma(\cdot)$ is a sigmoid function, and $\mathbf{h}_v$ and $\mathbf{h}_u$ are the final embeddings of nodes $v$ and $u$.

The model is trained using a binary cross-entropy loss function, which is commonly used for link prediction tasks. The loss function is defined as:
\begin{equation}
\mathcal{L} = -\frac{1}{|\mathcal{V}|} \sum_{(v,u) \in \mathcal{V}} \left[y_{vu} \log(s_{vu}) + (1 - y_{vu}) \log(1 - s_{vu})\right],
\end{equation}
where $\mathcal{V}$ is the set of all node pairs in the training data, $y_{vu} \in \{0, 1\}$ indicates whether a link exists between nodes $v$.

\section{Experiments}
In this section, we validate the effectiveness of the MGLP method through various experimental setups, addressing the following research questions: \textbf{RQ1:} How does the performance of the MGLP method compare to existing baseline methods? \textbf{RQ2:} How do the individual components of the MGLP method affect its overall performance? \textbf{RQ3:} How does the number of coarse-grained granular balls influence the performance of the MGLP method?

\begin{table}[t]
    \centering
    \begin{adjustbox}{width=1.1\textwidth, center}
        \begin{tabular}{|c|cccccc|}
            \hline
             Baselines & COLLAB & DDI & PubMed & Cora & Citeseer & Facebook  \\
             \hline
             Adamic Adar & 53.00 $\pm$ 0.00 & 18.61 $\pm$ 0.00 & 66.89 $\pm$ 0.00 & 77.22 $\pm$ 0.00 & 68.94 $\pm$ 0.00 & 99.41 $\pm$ 0.00 \\
             MF & 38.74 $\pm$ 0.30 & 17.92 $\pm$ 3.57 & 58.18 $\pm$ 0.01 & 51.14 $\pm$ 0.01 & 50.54 $\pm$ 0.01 & 98.80 $\pm$ 0.00 \\
             Node2Vec & 41.36 $\pm$ 0.69 & 21.95 $\pm$ 1.58 & 80.32 $\pm$ 0.29 & 84.49 $\pm$ 0.49 & 80.00 $\pm$ 0.68 & 86.49 $\pm$ 4.32 \\
             GCN(GAE) & 44.14 $\pm$ 1.45 & 37.07 $\pm$ 5.07 & 95.80 $\pm$ 0.13 & 88.68 $\pm$ 0.40 & 85.35 $\pm$ 0.60 & 98.66 $\pm$ 0.04 \\
             GCN(MLP) & 44.29 $\pm$ 1.88 & 39.31 $\pm$ 4.87 & 95.83 $\pm$ 0.80 & 90.25 $\pm$ 0.53 & 81.47 $\pm$ 1.40 & 99.43 $\pm$ 0.02 \\
             GraphSAGE & 48.62 $\pm$ 0.87 & 44.82 $\pm$ 7.32 & 96.58 $\pm$ 0.11 & 90.24 $\pm$ 0.34 & 87.37 $\pm$ 1.39 & 99.29  $\pm$ 0.01 \\
             GAT & 44.14 $\pm$ 5.95 & 29.53 $\pm$ 5.58 & 85.55 $\pm$ 0.23  & 82.59 $\pm$ 0.14 & 87.29 $\pm$ 0.11 & 99.37 $\pm$ 0.00 \\
             JKNet & 48.84 $\pm$ 0.83 & 57.98 $\pm$ 7.68 & 96.58 $\pm$ 0.23 & 89.05 $\pm$ 0.67 & 88.58 $\pm$ 1.78 & 99.43 $\pm$ 0.02 \\
             P-GNN & - & 1.14 $\pm$ 0.25 & 87.22 $\pm$ 0.51 & 85.92 $\pm$ 0.33 & 90.25 $\pm$ 0.42 & 93.13 $\pm$ 0.21 \\
             GTrans+LPE & 11.19 $\pm$ 0.42 & 9.22 $\pm$ 0.20 & 81.15 $\pm$ 0.12 & 79.31 $\pm$ 0.09 & 77.49 $\pm$ 0.02 & 99.27 $\pm$ 0.00 \\
             GCN+LPE & 49.75 $\pm$ 1.35 & 38.18 $\pm$ 7.62 & 95.50 $\pm$ 0.13  & 76.46 $\pm$ 0.15 & 78.29 $\pm$ 0.21 & 99.17 $\pm$ 0.00\\
             GCN+DE & 53.44 $\pm$ 0.29 & 26.63 $\pm$ 6.82 & 95.42 $\pm$ 0.08 & 89.51 $\pm$ 0.12 & 86.49 $\pm$ 0.11 & 99.38 $\pm$ 0.02 \\
             GCN+LRGA & 52.21 $\pm$ 0.72 & 62.30 $\pm$ 9.12 & 93.53 $\pm$ 0.25 & 88.83 $\pm$ 0.01 & 87.59 $\pm$ 0.03 & 99.42 $\pm$ 0.05 \\
             SEAL & 53.72 $\pm$ 0.95 & 26.25 $\pm$ 8.00 & 95.86 $\pm$ 0.28 & 92.55 $\pm$ 0.50 & 85.82 $\pm$ 0.44 & 99.60 $\pm$ 0.02 \\
             NBF-net & -  & 4.03 $\pm$ 1.32 & 97.30 $\pm$ 0.45 & 94.12 $\pm$ 0.17 & 92.30 $\pm$ 0.23 & 99.42 $\pm$ 0.04 \\
             PEG-DW+ & 53.70 $\pm$ 1.18 & 47.88 $\pm$ 4.56 & 97.21 $\pm$ 0.18 & 93.12 $\pm$ 0.12 & 94.18 $\pm$ 0.18 & 99.57 $\pm$ 0.05 \\
             {HPLC} & \underline{56.04 $\pm$ 0.28} & \underline{70.03 $\pm$ 7.02} & \underline{97.38 $\pm$ 0.34} & \underline{94.95 $\pm$ 0.18} & \underline{96.15 $\pm$ 0.19} & \textbf{99.69 $\pm$ 0.00} \\
             \textbf{MGLP} & \textbf{56.59 $\pm$ 0.53} & \textbf{71.01 $\pm$ 6.26} & \textbf{97.92 $\pm$ 0.23} & \textbf{95.80 $\pm$ 0.15} & \textbf{96.45 $\pm$ 0.17} & \underline{99.66 $\pm$ 0.00} \\
             \hline
        \end{tabular}
    \end{adjustbox}
    \caption{Comparison of experimental results between MGLP and baseline methods}
    \label{tab:comparison}
    \label{result}
\end{table}

\subsection{Experimental Setup}
\textbf{Datasets } We conducted experiments on six commonly used datasets for link prediction. Among them, Cora, Citeseer, Pubmed \cite{kipf2017semisupervised}, and Facebook are relatively small-scale datasets, while the remaining datasets consist of dense or large graphs provided by OGB \cite{hu2020open}. The detailed statistics of the datasets are shown in Table \ref{tab:datasets}.

\begin{table}[h!]
\centering

\setlength{\tabcolsep}{6pt}
\renewcommand{\arraystretch}{1.1} 
\begin{tabular}{@{}lrrrrrrr@{}}
    \toprule
    Dataset & \# Nodes & \# Edges & $\frac{\# \text{Edges}}{\# \text{Nodes}}$& Split ratio\\ 
    \midrule
    Cora        & 2,708    & 7,986      & 2.95    & 70/10/20 \\
    Citeseer    & 3,327    & 7,879      & 2.36    & 70/10/20 \\
    PubMed      & 19,717   & 64,041     & 3.25    & 70/10/20 \\
    Facebook    & 4,039    & 88,234     & 21.85   & 70/10/20 \\
    DDI         & 4,267    & 1,334,889  & 312.84  & 80/10/10 \\
    COLLAB      & 235,868  & 1,285,465  & 5.41    & 92/4/4   \\
    \bottomrule
\end{tabular}
\caption{Dataset statistics}
\label{tab:datasets}
\end{table}

\textbf{Evaluation metrics }  We follow previous work \cite{kim2024hierarchical}. For the Cora, Citeseer, Pubmed, and Facebook datasets, we used the area under the ROC curve (AUC) as the evaluation metric for the model. For the DDI and COLLAB datasets, we calculated the proportion of positive edges in the test data that ranked within the top-k positions to assess the model's performance. Additionally, we computed the average value of the evaluation metrics across different datasets to reflect the overall effectiveness of the MGLP method.

\textbf{Baselines }  To evaluate the effectiveness of the MGLP method on the link prediction task, we compared our approach with 17 existing baseline methods, including Adamic Adar (AA) \cite{adamic2003friends}, Matrix Factorization (MF) \cite{koren2009matrix}, Node2Vec \cite{grover2016node2vec}, GCN \cite{kipf2017semisupervised}, GraphSAGE \cite{hamilton2017inductive}, GAT \cite{velickovic2018graph}, P-GNN, NBF-Net \cite{zhu2021neural}, JKNet \cite{xu2018representation}, SEAL, GCN+DE \cite{li2020distance}, GCN+LPE \cite{dwivedi2020benchmarking}, GCN+LRGA \cite{puny2021global}, Graph Transformer+LPE \cite{dwivedi2021generalization}, PEG-DW+ \cite{wang2022equivariant} and HPLC.


\textbf{Implementation Details } All experiments in the paper were implemented using Python and PyTorch Geometric and conducted on an Intel(R) Xeon(R) W-2245 CPU @ 3.90GHz and an NVIDIA GeForce RTX 3090 GPU. Link prediction performance was evaluated based on the ranking of positive edges relative to negative edges in the test datasets. We uniformly integrated a GCN model with the following hyperparameter settings: a learning rate of 0.001, a hidden layer size of 256, a dropout rate of 0.2, binary cross-entropy (BCE) as the loss function, and Adam as the optimizer. For negative edges, we randomly selected an equal number of negative edges corresponding to the positive edges for the experiments.

\begin{table}[t!]
    \centering
    
    \resizebox{1\textwidth}{!}{
    \begin{tabular}{|c|c|c|c|c|c|c|}
        \hline 
       \textbf{MG} & \textbf{CV} & \textbf{DV} & \textbf{DDI} & \textbf{PubMed} & \textbf{Cora} & \textbf{Citeseer} \\
         \hline
         \ding{55} & \ding{55} & \ding{55}& 40.23 $\pm$ 5.14 & 95.80 $\pm$ 0.54 & 91.05 $\pm$ 0.44 & 82.55 $\pm$ 0.64 \\
         \hline
        \ding{55} & \ding{52} & \ding{55} & 64.91 $\pm$ 8.25 & 96.16 $\pm$ 0.19 & 92.66 $\pm$ 0.16 & 93.76 $\pm$ 0.16 \\
         \hline
          \ding{55} & \ding{55} & \ding{52}& 65.09 $\pm$ 4.64& 96.22 $\pm$ 0.14 & 93.08 $\pm$ 0.18 & 95.44 $\pm$ 0.23 \\
         \hline
          \ding{55} & \ding{52} & \ding{52}& 68.01 $\pm$ 7.23 & 96.61 $\pm$ 0.12 & 93.69 $\pm$ 0.23 & 95.00 $\pm$ 0.11\\
         \hline
         \ding{52} & \ding{55} & \ding{52}& 68.86 $\pm$ 5.76 & 96.76 $\pm$ 0.16 & 94.37 $\pm$ 0.17 & 95.57 $\pm$ 0.21 \\
         \hline
        \ding{52} & \ding{52} & \ding{55}& 66.28 $\pm$ 9.14 & 96.37 $\pm$ 0.15 & 94.01 $\pm$ 0.22 & 94.77 $\pm$ 0.15 \\
         \hline
         \ding{52} & \ding{52} & \ding{52}& \textbf{71.01 $\pm$ 6.26} & \textbf{97.92 $\pm$ 0.23} & \textbf{95.80 $\pm$ 0.15} & \textbf{96.45 $\pm$ 0.17} \\
         \hline
    \end{tabular}
    }
    \caption{Ablation study of MGLP}
    \label{tab:albation}
\end{table}


\begin{figure}[t]
\centering
    \includegraphics[width=1\textwidth]{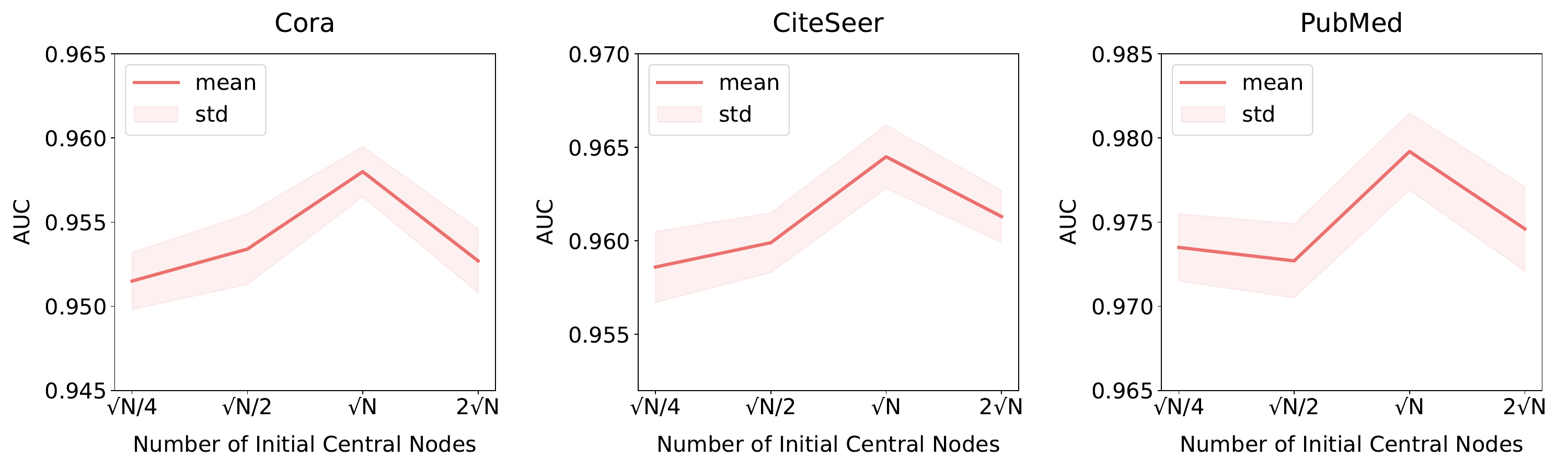}
    \caption{Parametric Analysis of Initial Granular-Ball Central Nodes }
    \label{fig:para_analys}
\end{figure}

\subsection{Overall Performance Comparison(RQ1)}
The experimental results of the MGLP compared with baselines are summarized in Table \ref{tab:comparison}. All methods were evaluated by performing 10 repeated experiments and calculating the average results. The best performance is highlighted in \textbf{bold}, while the second-best performance is \underline{underlined}. “-” indicates "Out of Memory" (OOM). 

From the table, we can derive the following observations: 1) \textbf{Outstanding performance of MGLP:} As quantified in Table \ref{tab:comparison}, MGLP consistently surpasses all 17 baselines across datasets, establishing state-of-the-art (SOTA) results. On the dataset DDI, MGLP achieves 71.01 ± 6.26 HITS@K, outperforming the previous SOTA (HPLC: 70.03 ± 7.02) by a 1.4 absolute gain. For PubMed, a sparse citation network, MGLP attains 97.92 ± 0.23 AUC versus HPLC’s 97.38 ± 0.34, demonstrating superior robustness to structural sparsity. On the massive COLLAB graph, MGLP achieves 56.59 ± 0.53 HITS@K—a 0.55 improvement over HPLC (56.04 ± 0.28), validating scalability. Cora and Citeseer further confirm MGLP’s dominance, exceeding topology-aware baselines like SEAL (92.55) and HPLC (94.95). This improvement stems from MGLP's ability to adaptively generate structures with varying granularities, allowing it to flexibly accommodate different data distribution characteristics, thereby enhancing model performance. 2) \textbf{Effective multi-granularity homogeneous structures and hierarchical relationships:} MGHD accurately captures the structural relationships in the original graph while effectively representing the hierarchical relationships in the multi-granularity tree. By integrating these two types of relationships, MGHD provides a more comprehensive distance representation in complex networks, thereby improving model performance.

\subsection{Ablation Experiment(RQ2)}
In this section, we conduct ablation studies on the components of the MGLP model, with the detailed results presented in Table~\ref{tab:albation}. In this table, "MG" "CV" and "DV" represent the use of multi-granularity information, center vectors, and distance vectors, respectively. We incorporate multi-granularity by retaining granular-ball information throughout the splitting process.

The experimental results demonstrate that these components contribute significantly to enhancing the model's performance: (1) When the multi-granularity(MG) module is removed, the performance of MGLP significantly degrades. This change highlights the importance of the multi-granularity feature, which enables the capture of both local and global structural characteristics, thereby providing a more comprehensive data representation. (2) The model performance also significantly declines when the center vectors(CV) generated by the HCG construction are not used. This indicates that refining the relationships between the center points at each hierarchical level in the HCG is beneficial for improving model performance. Refining the interrelationships between the center points helps form more effective node embeddings. (3) The removal of the distance vectors(DV) generated by MGHD also leads to a significant decline in performance. This proves that the proposed distance computation paradigm, integrates structural relationships from the original graph with hierarchical relationships in the multi-granularity tree, offering more precise and effective distance representations. Such representations enhance the model's generalization capability in complex graph structures, leading to superior experimental performance. 

\subsection{Parameter Analysis(RQ3)}
In this part, we conducted experiments with varying initial numbers of granular-ball central nodes, as illustrated in Figure \ref{fig:para_analys}. Experimental results show that when we choose $\mathcal{\sqrt{N}}$ granular-ball central nodes, the performance is the best. This is because the choice of $\mathcal{\sqrt{N}}$ provides a balanced granularity level, ensuring that structures of different granularities can effectively capture local features while retaining the ability to abstract global relationships. This also serves as a validation of previous work \cite{xie2020new}.

\section*{Conclusion}
In this paper, we propose an adaptive multi-granularity position embedding method via granular-ball for link prediction, MGLP. MGLP adaptively refines the graph into structures of varying levels of granularity, facilitating the effective representation of homogeneous structures across multi-granularity. And we introduce the HCG to encode the relative positions of the central nodes, serving as an additional representation of the node positional information. Furthermore, we propose a novel distance encoding mechanism MGHD that integrates information from homogeneous structures and multi-granularity relationships within a hierarchical tree framework. Experimental results on six datasets demonstrate the superiority of the MGLP framework.

\bibliographystyle{elsarticle-harv} 
\bibliography{KBS2025}











\end{document}